\documentclass{aa}
\usepackage{graphicx}
\usepackage{siunitx}
\usepackage[varg]{txfonts}

\usepackage{natbib,twoopt}
\usepackage{xcolor}
\usepackage[breaklinks=true, colorlinks=true, linkcolor=blue, citecolor=blue]{hyperref}

\include{newcom}

\begin{document}
\title{eROSITA Spectro-Imaging Analysis of the Abell 3408 Galaxy Cluster}

\titlerunning{eROSITA Spectro-Imaging Analysis of the Abell 3408 Galaxy Cluster}

\author{J. Iljenkarevic \inst{1}\and T. H. Reiprich \inst{1} \and F. Pacaud \inst{1} \and A. Veronica \inst{1} \and B. Whelan \inst{1} \and J. Aschersleben \inst{1} \and K. Migkas \inst{1}  \and E. Bulbul \inst{2} \and J. S. Sanders \inst{2} \and M. E. Ramos-Ceja \inst{2} \and  T. Liu \inst{2} \and V. Ghirardini \inst{2} \and  A. Liu \inst{2} \and T. Boller \inst{2}}

\institute{Argelander-Institut f\"ur Astronomie (AIfA), Universit\"at Bonn,
              Auf dem H\"ugel 71, 53121 Bonn, Germany\\
         \and
             Max-Planck-Institut f\"ur extraterrestrische Physik, Giessenbachstra{\ss}e 1, 85748 Garching, Germany}
\date{Received \dots; Accepted \dots}

\abstract
{The X-ray telescope eROSITA onboard the newly launched Spectrum-Roentgen-Gamma (SRG) mission serendipitously observed the galaxy cluster Abell 3408 (A3408) during the Performance Verification observation of the AGN 1H0707-495.
eROSITA's one degree field-of-view allowed us to trace the intriguing elongated morphology of the nearby ($z=0.0420$) A3408 cluster. Despite its brightness ($F_{500}\approx 7\times 10^{-12}$\,ergs/s/cm$^2$) and large extent ($r_{200} \approx 21'$),
it has not been observed by any modern X-ray observatory in more than 20 years. A neighbouring cluster in NW direction, A3407 ($r_{200} \approx 18'$, $z=0.0428$), appears to be close at least in projection ($\sim 1.7 \:{\rm Mpc}$).
Potentially, this cluster pair could be in a pre- or post-merger state.} 
{We aim to determine detailed thermodynamical properties of this special cluster system for the first time.
Furthermore, we aim to determine which of the previously suggested merger scenarios (pre- or post-merger) is preferred.}
{We perform a detailed X-ray spectro-imaging analysis of A3408. We construct particle background subtracted and exposure corrected images and surface brightness profiles in different sectors. The spectral analysis is performed out to $1.4r_{500}$ and includes normalization, temperature and metallicity profiles determined from elliptical annuli aligned to the elongation of A3408. Additionally, a temperature map is presented depicting the
distribution of the ICM temperature. Furthermore, we make use of data from the ROSAT All Sky Survey to estimate some bulk properties of A3408 and A3407, using the growth curve analysis method and scaling relations.} 
{The imaging analysis shows a complex morphology of A3408 with a strong elongation in SE-NW direction. This is quantified by comparing the surface brightness profiles of the NW, SW, SE and NE directions, where the NW and SE directions show a significantly higher surface brightness compared to the other directions. 
We determine a gas temperature ${\rm k_B}T_{500}=(2.23\pm0.09)\:{\rm keV}$ in the range $0.2r_{500}$ to $0.5r_{500}$ from the spectral analysis. The temperature profile reveals a hot core within two arcmin of the emission peak, ${\rm k_B}T=3.04^{+0.29}_{-0.25}\:{\rm keV}$.
Employing a mass--temperature relation, we obtain $M_{500} = (9.27 \pm 0.75)\times 10^{13}M_{\odot}$
iteratively.
The $r_{200}$ of A3407 and A3408 are found to overlap in projection which makes ongoing interactions plausible. The two-dimensional temperature map reveals higher temperatures in W than in E direction.
}
{The elliptical morphology together with the temperature distribution suggests that A3408 is an unrelaxed system.
The system A3407 and A3408 is likely in a pre-merger state with some interactions already affecting the ICM thermodynamical properties. In particular, increased temperatures in the direction of A3407 indicate adiabatic compression or shocks due to the starting interaction.}
\maketitle

\section{Introduction}
\label{sec:1}
Galaxy clusters are the largest gravitationally bound systems in the Universe located at the nodes of the cosmic filaments. They form by merging processes of systems with smaller masses and grow by accretion of matter  along the filaments. The interactions between galaxy clusters are complex, i.e. shock fronts caused by infalling subclusters or cold fronts with different origins can occur in the intracluster medium (ICM) as evidenced by X-ray spectro-imaging analyses \citep[e.g.,][]{Markevitch_2007}.
Studying galaxy clusters and their interactions offers an insight into the large scale structure of the Universe and its evolution. Bright, low-redshift, potentially interacting groups and clusters such as A3408 \citep[$z=0.0420$,][]{Struble_1999} and A3407 \citep[$z=0.0428$,][]{Struble_1999}
provide the opportunity to widen this knowledge. 

eROSITA (extended ROentgen Survey with an Imaging Telescope Array) is a newly launched X-ray telescope \citep{Predehl_2021}. It is the primary instrument of the German-Russian Spectrum-Roentgen-Gamma (SRG) mission \citep{2021arXiv210413267S}. 
SRG/eROSITA was launched on 13th of July 2019 from Baikonur in Kazakhstan and is performing a deep survey of the entire X-ray sky since 8th of December 2019. The main scientific purpose of the all-sky survey is to study galaxy clusters and their cosmic evolution \citep{merloni2012erosita}. Before the all-sky survey started, eROSITA went through a Calibration and Performance Verification (CAL-PV) phase, in which the instrument carried out a series of observations on a variety of targets.

The galaxy cluster Abell 3408 (A3408) has been serendipitously observed during the PV observation of the Active Galactic Nuclues (AGN) 1H0707-495 \citep{Boller_2021}. eROSITA's one degree field-of-view allowed us to trace the unique elongated morphology of A3408. Despite its brightness and large extent, it has not been observed by any modern X-ray observatory in the last 20 years. 
The discovery of an arc-like object ($z=0.073$) in the optical band attracted attention to A3408 \citep{Campusano_1998, Cypriano_2001}, as it was interpreted as an image of a background galaxy that is distorted due to the strong gravitational lensing effect. The latest ICM temperature measurement was reported in \cite{Katayama_2001}, who measured a temperature of ${\rm k_B}T=(2.9 \pm 0.2) \:{\rm keV}$ within a radius of $800\:{\rm kpc}$, utilizing data from the ASCA satellite.
\cite{Chow_Martinez_2014} classified A3408 and A3407 as a Supercluster based on the small distance between the two neighbouring clusters.
Furthermore, there seems to be an indication that A3408 is interacting with 
A3407 \citep{Nascimento_2016}. Based on a dynamical study of 122 member galaxies, Nascimento et al. conclude that A3408 and A3407 could be in a pre-merger scenario crossing each other in less than $\sim 1 h^{-1}$~Gyr or in a post-merger scenario where they crossed each other $\sim 1.65 h^{-1}$~Gyr ago. 

This article presents a detailed spectro-imaging analysis of A3408 with the main goal to describe the unique cluster morphology, temperature and abundance distribution. Furthermore, we search for indications for possible interactions between A3408 and A3407. 
The article is structured as follows: Sec.~\ref{sec:6} contains the analysis of the ROSAT data of A3408 and A3407. Sec.~\ref{sec:2} and Sec.~\ref{sec:3} describe the data processing of the eROSITA data and source detection, respectively. The eROSITA imaging analysis including results is presented in Sec.~\ref{sec:4} and the spectral analysis in Sec.~\ref{sec:7}. The results are discussed in Sec.~\ref{sec:8}. Finally, Sec.~\ref{sec:9} contains the conclusions and an outlook on further work. The applied cosmological model is a flat $\Lambda$CDM cosmology with $\Omega_{\rm m} = 0.3$, $\Omega_{\Lambda} = 0.7$ and ${\rm H}_0 = 70\:{\rm km/s/Mpc}$ with ${\rm H}_0 = 100h\:{\rm km/s/Mpc}$ and physical angular scales of $0.8289 \:{\rm kpc/arcsec}$ at $z=0.0420$ and $0.8439 \:{\rm kpc/arcsec}$ at $z=0.0428$.

\section{ROSAT Cluster Properties}
\label{sec:6}

Both A3407 and A3408 show significant emission in the ROSAT All Sky Survey maps. In order to get some insights on the properties of A3407, we estimated its properties from the hard band images of the survey (0.5-2 keV), and applied the same methods to A3408 for consistency.

The process relies on the growth curve analysis method developed by \citep{Bohringer_2001} and our implementation is described in detail in \cite{Xu_2018}. It first estimates, for each cluster, a local background from a large annulus around its center ($25'-45'$). Then, it extracts an integrated, background subtracted, count-rate profile out to $25'$, in bins of $0.5'$. In determining both the background and growth curve, we masked contaminating sources and corrected for the missing flux assuming azimuthal symmetry. At this stage, we measure a reference source count-rate $CR_\mathrm{ap}$ within the significance radius, $\theta_\mathrm{ap}$, defined as the aperture out of which an increase of the total source count-rate can no longer be detected. In a second part of the process, we seek to estimate cluster properties within, $r_\mathrm{500}$, the radius within which the average cluster density is 500 times larger than the critical density of the Universe at the cluster's redshift $z$. For this, we relied on a set of scaling relations connecting the cluster mass $M_{500}$ to its luminosity $L_\mathrm{500}$ in 0.1-2.4 keV \citep{Schellenberger_2017a} and temperature $T_\mathrm{X}$ \citep{Lovisari_2015}, in order to derive the ROSAT count-rate $CR_\mathrm{500}$ expected within $r_\mathrm{500}$ as a function of $M_\mathrm{500}$ and $z$. Using an iterative procedure, the algorithm identifies the unique set of cluster parameters whose predicted ($r_\mathrm{500}$,$CR_\mathrm{500}$) correspond to the cluster photometric measurements ($\theta_\mathrm{ap}$,$CR_\mathrm{ap}$). The latter conversion assumes a $\beta$-model surface brightness with $\beta=2/3$ and $r_\mathrm{c}=0.12\times r_\mathrm{500}$ and accounts for the redistribution arising from the ROSAT PSF.

\cite{Reiprich_2013} find that $r_{\rm vir} (z=0) \approx r_{100} \approx 1.36 r_{200}$ and $r_{500} \approx 0.65 r_{200}$. Using these relations, we can roughly estimate $r_{200}$ and $r_{100}$ of A3408 and A3407 respectively.  

The results of the analysis of the ROSAT data of both clusters is summarized in Tab.~\ref{tbl:3}. 
All radii are illustrated in Sec.~\ref{sec:7}, where the center of A3407 is taken from the catalog in \cite{Abell_1989}.
The figure shows that the $r_{100}$ and $r_{200}$ of both clusters overlap at least in projection, i.e. interactions are plausible. The distance between the cluster centers at redshift $z=0.042$ is $\sim 1.7 \: {\rm Mpc}$. 

\begin{table*}[h!]
	\centering
	\begin{tabular}[c]{ccccccccccc}
		\hline
	     ~&&~& $F_{500}$ & $L_{500}$ & ${\rm k_B}T$ & $M_{500}$ & $r_{500}$ & $r_{200}$ & $r_{100}$ &~\\
		 ~&&~& [$10^{-12}~{\rm ergs/s/cm^2}$] & [$10^{43}~{\rm ergs/s}$]& [keV] & [$10^{13}M_{\odot}$] & [kpc] & [kpc] & [kpc] &~\\
		\hline\hline
		&&&&&&& \\
        ~& A3408  &~& $ 6.78 \pm 1.24$ & $2.76 \pm 0.50$ & $ 2.30 \pm 0.18$ & $9.84 \pm 1.34$ & $693 \pm 31$ & $1066 \pm 48$ & $1450 \pm 66$ &~\\
        ~& A3407  &~& $ 3.23 \pm 0.72 $ & $1.37 \pm 0.30$ & $ 1.69 \pm 0.16$ & $5.83 \pm 0.97$ & $581 \pm 32$ & $895 \pm 49$ & $1217 \pm 67$ &~\\
        &&&&&&& \\
		\hline
	\end{tabular}
	\caption{Cluster properties 
	determined from ROSAT data and scaling relations as described in Sec.~\ref{sec:6}.
	}
	\label{tbl:3}
\end{table*}

\section{Data Processing}
\label{sec:2}

The AGN 1H0707-495 was observed by eROSITA on 11th of October 2019 \citep{Boller_2021}. The corresponding Observation ID is 300003. 
This pointed PV-observation was performed by the Telescope Modules (TM) 1, 2, 5, 6 and 7. The observations by TM 1 and 2 have not reached to the full exposure time of 50 ks. TM 3 and 4 were not active during the entire observation.
All observation times are summarized in Tab.~\ref{tbl:1}.

\begin{table}[h!]
	\centering
	\begin{tabular}[c]{cccc}
		\hline
		ObsID & TM & Observing time & SRG Observing mode\\
		\hline\hline
		300003 & 1  & 14.2 ks & Pointing\\		
        300003 & 2 & 10.3 ks & Pointing\\
        300003 & 5 & 49.4 ks & Pointing\\
        300003 & 6  & 50.3 ks & Pointing\\
        300003 & 7 & 49.8 ks & Pointing\\
		\hline
	\end{tabular}
	\caption{PV-observation (ObsID: 300003) of the AGN 1H0707-495 and the A3408 galaxy cluster on 11th of October 2019 by eROSITA.
	}
	\label{tbl:1}
\end{table}

\subsection{Data preparation}
\label{sec:2.1}
The data used in this work is obtained from the processing configuration 001.
The data reduction is performed using the eROSITA Science Analysis Software System (eSASS, Brunner et al., in prep.), release 201009. 

The event files are generated in the full energy band of $0.2 - 10.0 \:{\rm keV}$ using the eSASS task \texttt{evtool} with parameters \texttt{pattern=15} and \texttt{flag=0xc00fff30}, i.e. all single, double, triple, and quadruple patterns are selected and events located in the strongly vignetted corners of the square CCDs are removed because of remaining uncertainties in the vignetting and PSF calibration.

In general, the observation could be affected by a temporally varying background component, e.g., by soft proton flares as observed by XMM-Newton \citep[e.g.,][]{Reiprich_2004}.
In order to check the quality of the data, light curves are created with the eSASS task \texttt{flaregti} in the $6.0 -  9.0 \:{\rm keV}$ energy band and inspected for each TM. The presence of a flare was determined for TM 5 and 7. Consequently, a consistent filtering procedure was applied to all event files. 
The filtering procedure includes automatic and additional filtering. The automatic filtering results in Good-Time Intervals (GTIs) from the standard eSASS processing. 
The additional filtering determines the $3\sigma$ interval by fitting a Gaussian fit to the histogram of the rates. Any interval beyond $\pm 3 \sigma$ is rejected from the data. Finally, the GTIs from both, automatic and additional filtering, are combined. These clean and filtered event files for each TM are used for the analysis in this work.
More information on the filtering procedure and all lightcurves in the $6.0 -  9.0 \:{\rm keV}$ energy band are shown in Appendix \ref{A1}.

\subsection{Light Leak}
\label{sec:2.2}

During the camera commissioning period, light leak contamination was detected in TM 5 and 7, i.e. parts of the respective CCDs are contaminated by optical photons.
These optical photons produce apparent X-ray events with energies below $ \sim 0.8 \: {\rm keV}$.
TM 1-4 and 6 are equipped with on-chip optical light filters and are therefore not affected by the light leak \citep[more details in][]{Predehl_2021}.
For this reason, the imaging and spectral analysis in this work applies different energy bands for TM with on-chip filters, i.e. TM 1, 2 and 6, in contrast to TM without on-chip filters, i.e. TM 5 and 7. 

\section{Source Detection}
\label{sec:3}

The eSASS source detection algorithm (Brunner et al., in prep.) detects point sources in the observation based on the sliding-cell method and is performed on combined clean and filtered event files of TM 1, 2, 5, 6 and 7.

First, the eSASS task \texttt{erbox} in {\it local mode} searches for signals that exceed a certain threshold and creates a catalog with detected sources. The minimum detection likelihood value applied here is $6.0$. In a next step, a background map is generated with the task \texttt{erbackmap} by masking out the previously detected sources and utilizing an adaptive smoothing algorithm. Finally, \texttt{erbox} performs a second source detection in {\it map mode} with respect to the newly created background map.

Afterwards, a maximum likelihood fitting method is applied on the new source catalog using the eSASS task \texttt{ermldet}. Each detected source is characterized and corresponding source parameters, such as detection likelihood and extent, are determined. The Point Spread Function (PSF) of the telescopes is taken into account in this step. 
The resulting catalog is then visually inspected to check that no point-sources are missed.
Sources in the final catalog  are excluded from the further analyses.

\section{eROSITA Imaging Analysis}
\label{sec:4}

The imaging analysis provides a first insight about the spatial distribution of the ICM emission of A3408. 

\subsection{Preparation}
The particle background subtraction and the exposure correction are accomplished following the procedure described in \cite{Reiprich_2020}. With this method an image is created in the $0.3 - 2.3 \: {\rm keV}$ energy band for TM 1, 2 and 6 and in the $1.0 - 2.3 \:{\rm keV}$ energy band for TM 5 and 7, in order not to be affected by the light leak. In the following, the combination of the two TM-dependent energy bands is referred to as ``soft band''. 
The image extraction for each TM is performed with the eSASS task \texttt{evtool}, while the corresponding vignetted and un-vignetted exposure maps are generated with \texttt{expmap}.

In this procedure the particle background is modelled using the eROSITA Filter-Wheel-Closed (FWC) observation data. 
The spatial variation of the particle background was found to be very small and temporally stable \citep{Freyberg_2020}. 
Within the procedure, a particle background map is created for each TM using the signal in the hard band ($6.0-9.0 \:{\rm keV}$) as scaling. The spatial distribution is provided by multiplying to the respective 
unvignetted exposure maps, which are normalized to unity by dividing each pixel by the sum of all the pixel values. These background maps are then subtracted from the respective individual photon images. 
The resulting particle background-subtracted photon images for the individual TMs are added up, resulting in a combined particle background-subtracted soft band photon image.

\begin{figure}[h!]
\centering
	\includegraphics[width=1.0\columnwidth]{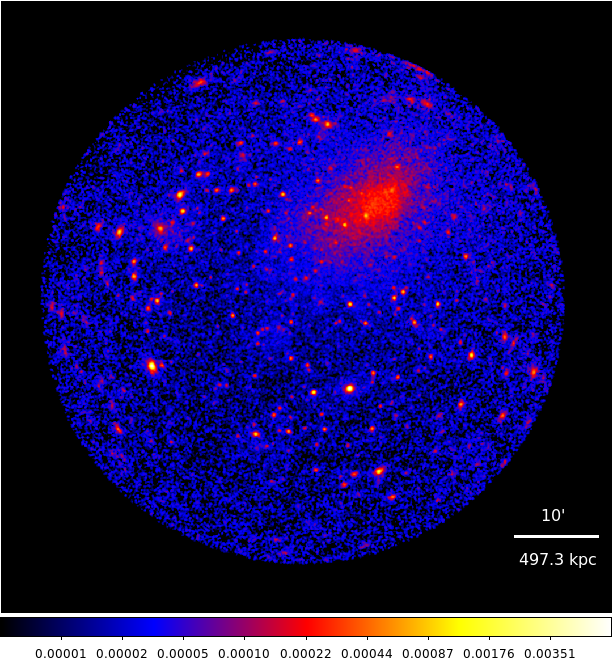}
	\caption{Particle background-subtracted and exposure-corrected photon image of A3408 with TM 1, 2, 5, 6 and 7 in the soft band ($0.3-2.3 \:{\rm keV}$ energy band for TM 1, 2 and 6 and $1.0-2.3 \:{\rm keV}$ energy band for TM 5 and 7) in logarithmic scale.
		\label{fig:1}}
\end{figure}

For the exposure correction, exposure maps are generated using \texttt{expmap} for each TM. The vignetting function is applied during this process. 
The exposure maps from TM 1, 2, and 6 are co-added, while those of the TMs affected by the light leak, i.e. TM 5 and 7, are co-added separately. Then both co-added exposure maps are combined with proper scaling, as described in detail in \cite{Reiprich_2020}.
In a last step, the particle background-subtracted image in the soft band is divided by the combined  exposure map. The final particle background-subtracted and exposure- and vignetting-corrected soft band image is shown in Fig.~\ref{fig:1}.

\subsection{Cluster Center and General Features}
\label{sec:4.2}

We apply an adaptive smoothing to the final soft band image after refilling removed point sources across the FoV with the surrounding background surface brightness. For this, we use the XMM-Newton Science Analysis System (SAS, 18.0.0) task \texttt{asmooth}\footnote{XMM-Newton SAS documentation, \texttt{asmooth} description: \url{https://xmm-tools.cosmos.esa.int/external/sas/current/doc/asmooth/index.html}} with the parameter \texttt{smoothstyle}~$=$~\texttt{adaptive}, further specifying a SNR of 10, a minimum number of 10 pixels and a maximum number of 20 pixels.

\begin{figure*}[h!]
\centering
	\includegraphics[width=2.0\columnwidth]{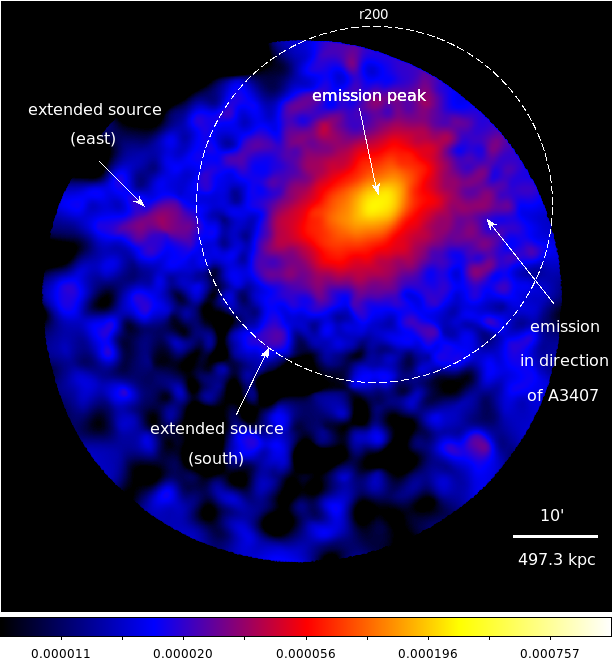}
	\caption{Adaptively smoothed particle background-subtracted and exposure-corrected photon image in soft band ($0.3-2.3 \:{\rm keV}$ energy band for TM 1, 2 and 6 and $1.0-2.3 \:{\rm keV}$ energy band for TM 5 and 7) in logarithmic scale.
	There seems to be diffuse emission out to beyond $r_{200}$ (white dashed line) in west direction, including possibly infalling galaxy groups (``extended source'') in east at similar redshift as A3408 and south (no redshift estimate available).
		\label{fig:2}}
\end{figure*}

Fig.~\ref{fig:2} shows the resulting adaptively smoothed soft band image of A3408. 
The image outlines the bright central region of A3408 and the elliptical appearance to outer radii. The center is shifted north-west relative to the larger scale emission, where the emission peak is determined to be at ${\rm RA}=107.0948 \:{\rm deg}$, ${\rm Dec}=-49.1933 \:{\rm deg}$. This peak is used as cluster center for all subsequent profile analyses. Furthermore, one can see enhanced emission in the west direction towards A3407 in the outermost regions. Moreover, there are some extended emission regions, e.g. in the east and the south of A3408. The extended source to the east is probably a galaxy group infalling into the main system. The brightest cluster galaxy is identified as 2MASS J07105737-4914161 at a redshift of $z=0.039$ \citep{2MASS_2012}. The general emission towards the east extends to beyond $r_{200}$. The extended source to the south may also be an infalling clump but no redshift estimate is available so a projected background cluster can presently not be excluded. However, there are three galaxies with redshift estimates within $10'$, all located at similar redshifts compared to A3408 which is an indication that the southern extended source is indeed an infalling clump.

\subsection{Surface Brightness Profile}
\label{sec:4.3}

The surface brightness profile enables us to study the morphology quantitatively by comparing different sectors.
We define the surface brightness as the number of photons per detector area (which corresponds to the effective area of the combined telescope, filter, detector system) per exposure time per area on the sky.
The calculation of photon counts in a given region is performed with \texttt{funcnts} from FUNTOOLS (1.4.7) on the particle background subtracted photon image in soft band.

\begin{figure*}[h!]
\centering
	\includegraphics[width=1.0\textwidth]{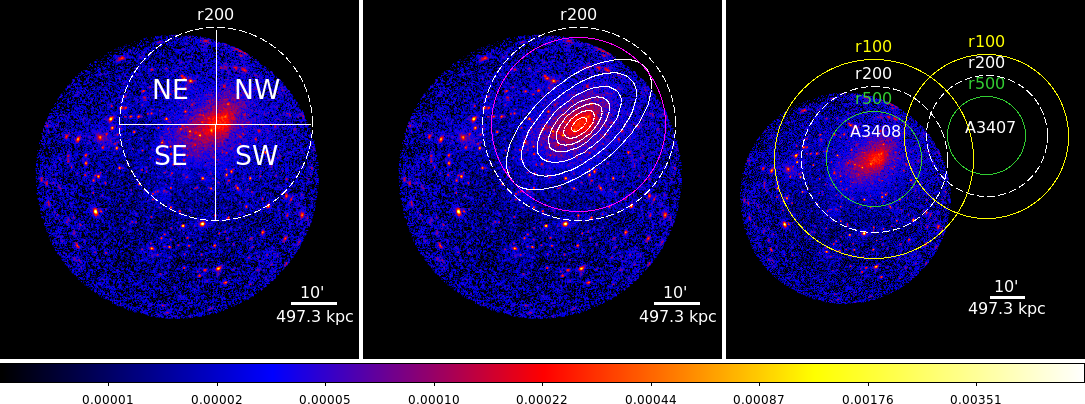}
	\caption{Particle background subtracted and exposure corrected photon image of TM 1, 2, 5, 6 and 7 in the
	soft band in logarithmic scale. 
	\textit{Left}: Four sectors, south-west, south-east, north-west and north-east out to $r_{200}$ (white, dashed), are illustrated. These areas are used to extract the counts for the surface brightness profiles respectively.
	\textit{Center}: Illustrated are elliptical annuli (white) out to a semi-major axis of $19'$ and a semi-minor axis of $9.5'$. The annuli have same eccentricity and angle of $40\:{\rm deg}$, chosen visually aligned to the cluster shape. These areas are used to obtain temperature, metallicity and normalization profiles. Furthermore, the temperature map was determined within a circular area of $19'$ (magenta) to ensure enough counts in the source regions for spectral fitting. 
	\textit{Right}: Shown are calculated characteristic radii $r_{500}$ (green), $r_{200}$ (white, dashed) and $r_{100}$ (yellow) of A3408 and A3407 from analyzing ROSAT data and applying the relations $r_{100} \approx 1.36 r_{200}$ and $r_{500} \approx 0.65 r_{200}$ in \cite{Reiprich_2013}.
		\label{fig:4}}
\end{figure*}

\begin{figure}
\centering
	\includegraphics[width=1.0\columnwidth]{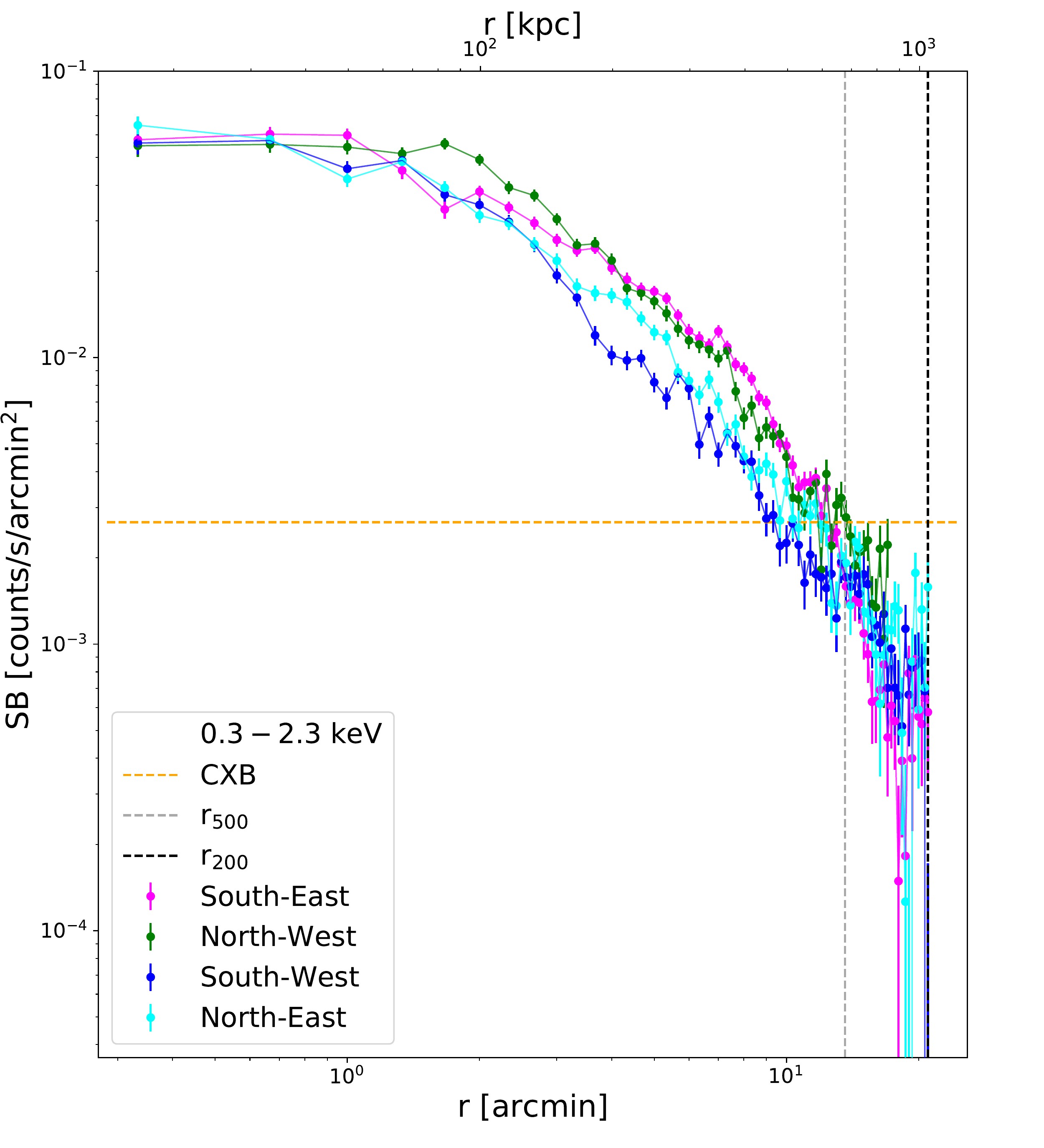}
	\caption{Surface brightness profiles
	obtained in four different sectors (south-west, south-east, north-west and north-east; as illustrated in Fig.~\ref{fig:4}, left). The CXB level is shown as orange dashed line, $r_{500}$ and $r_{200}$ are shown as grey and black dashed lines, respectively.
		\label{fig:3}}
\end{figure}

The surface brightness profile is calculated for four different sectors, north-west (NW), south-west (SW), south-east (SE), north-east (NE), as shown in Fig.~\ref{fig:4}, left, where the profile is centered at the cluster center determined in Sec.~\ref{sec:4.2} and most emission can be seen in SE-NW along the elongation of A3408.
In  detail, annuli with a width of $20''$ and a maximum outer radius of $r_{200}$ are chosen to sum up the counts from the particle background-subtracted image, as well as to determine the average from the exposure map. The surface brightness profile of the NW sector only contains bins that are at least 50\% inside the FoV. In a next step, the source counts are divided by the average exposure in a given source area. Additionally, the X-ray fore- and background (CXB) counts are obtained from the area of $r_{100}$ to the edge of the FoV and subtracted from the source counts after exposure and area scaling. 
We obtain the error of each source region from \texttt{funcnts} 
(using the particle background subtracted photon image). Then we also obtain error from Sky background region. The two errors (also from the exposure) are propagated to calculate the surface brightness profile.
The resulting surface brightness profile for each sector is shown in Fig.~\ref{fig:3}. Comparing all profiles, one can see enhanced surface brightness in NW in the central regions of $\sim 1'$ to $\sim 3'$. From there on the surface brightness is increasing in SE relative to the other directions, i.e. it confirms the elongation of A3408 in SE-NW direction. At outer radii ($>10'$), all profiles become more consistent with each other, while there is still enhanced emission in NW as previously seen in the adaptively smoothed image. Out to $r_{200}$ there is slightly more emission in NE compared to SE.

\section{eROSITA Spectral Analysis}
\label{sec:7}

The extraction procedure of all spectra and response files is performed using the eSASS task \texttt{srctool}. 
It creates the spectrum files for selected source and background regions and it also extracts the corresponding response matrices (RMFs) and  effective  areas (ARFs).
The point sources detected in the image (see Sec.~\ref{sec:3}) are excluded from source and background regions throughout the spectral analysis.

The X-ray spectral fitting package XSPEC 12.10.1 is used for spectral fitting \citep{XSPEC} applying $cstat$ likelihood for Poission statistics. The solar abundance table is taken from \cite{Asplund_2009}. 
The source spectra of all TM are modelled simultaneously with the background spectra and the instrumental background obtained from the FWC data of all TM. 
The $0.3-8.0 \:{\rm keV}$ energy band is used for TM 1, 2 and 6, while the lower energy used in TM 5 and 7 is $1.0 \:{\rm keV}$ to avoid the effects of optical light leak.

The cluster emission is modelled by an absorbed thermal  \texttt{apec} model with ATOMDB version 3.0.9 \citep{Foster_2012} within XSPEC and the absorption model TBabs for Galactic absorption \citep{Wilms_2000}. 
In the thermal \texttt{apec} model, the redshift $z$ is frozen to 0.0420 \citep{Struble_1999}, while cluster temperature $T$, metallicity $Z$ and normalization $K$ are free to vary. The normalization $K$ is defined as 

\begin{align}
    K = \frac{10^{-14}}{4~\pi~[D_A(z+1)]^2} \int n_e n_H ~dV.
\end{align}

$D_A$ is the angular diameter distance in units of cm, $n_e$ is the electron density and $n_H$ is the hydrogen density in units of cm$^{-3}$.
For the absorption model TBabs, we set the equivalent hydrogen column density $N_H$ to $6 \times 10^{20} \:{\rm ~atoms~cm}^{-2}$ \citep{Willingale_2013}.

The observed background usually consists of the Particle background and Cosmic X-ray Background (CXB).
The particle background is produced by energetic particles which interact with the surrounding material or the detector itself.
The CXB is composed of faint X-ray point sources that are unresolved by eROSITA.
Furthermore, the Local Hot Bubble (LHB) and the Galactic Halo (GH) also contribute to the CXB \citep{Lumb_2002, Kuntz_2008, Gilli_2006}. 

The particle background is estimated with eROSITA FWC observation data. 
The corresponding model contains Gaussion emission lines and powerlaw models with free normalizations.

The background spectra for the CXB model are extracted from a region starting at $r_{200}$ of the cluster out to the edge of the FoV excluding point sources. 
The CXB model contains three components: An unabsorbed thermal \texttt{apec} model with a temperature of $0.099 \:{\rm keV}$ for LHB (metallicity set to 1.0, redshift set to 0.0), an absorbed thermal model with a temperature of $0.225 \:{\rm keV}$ \citep{McCammon_2002} for GH (metallicity set to 1.0, redshift frozen to 0.0) and an absorbed powerlaw with a photon index of 1.41 \citep{DeLuca_2004} for the unresolved point sources. In addition, we add an absorbed thermal \texttt{apec} model to account for residual cluster emission in the background region. 

\subsection{eROSITA Cluster Properties}
\label{sec:7.2}

Cluster masses are not directly measurable. One way to determine them is through the use of scaling relations using X-ray observables. The $M_{500}-T$ relation enables the calculation of the cluster mass $M_{500}$ by the ICM temperature $T$. 

\begin{figure}
	\includegraphics[width=0.48\columnwidth,angle=270]{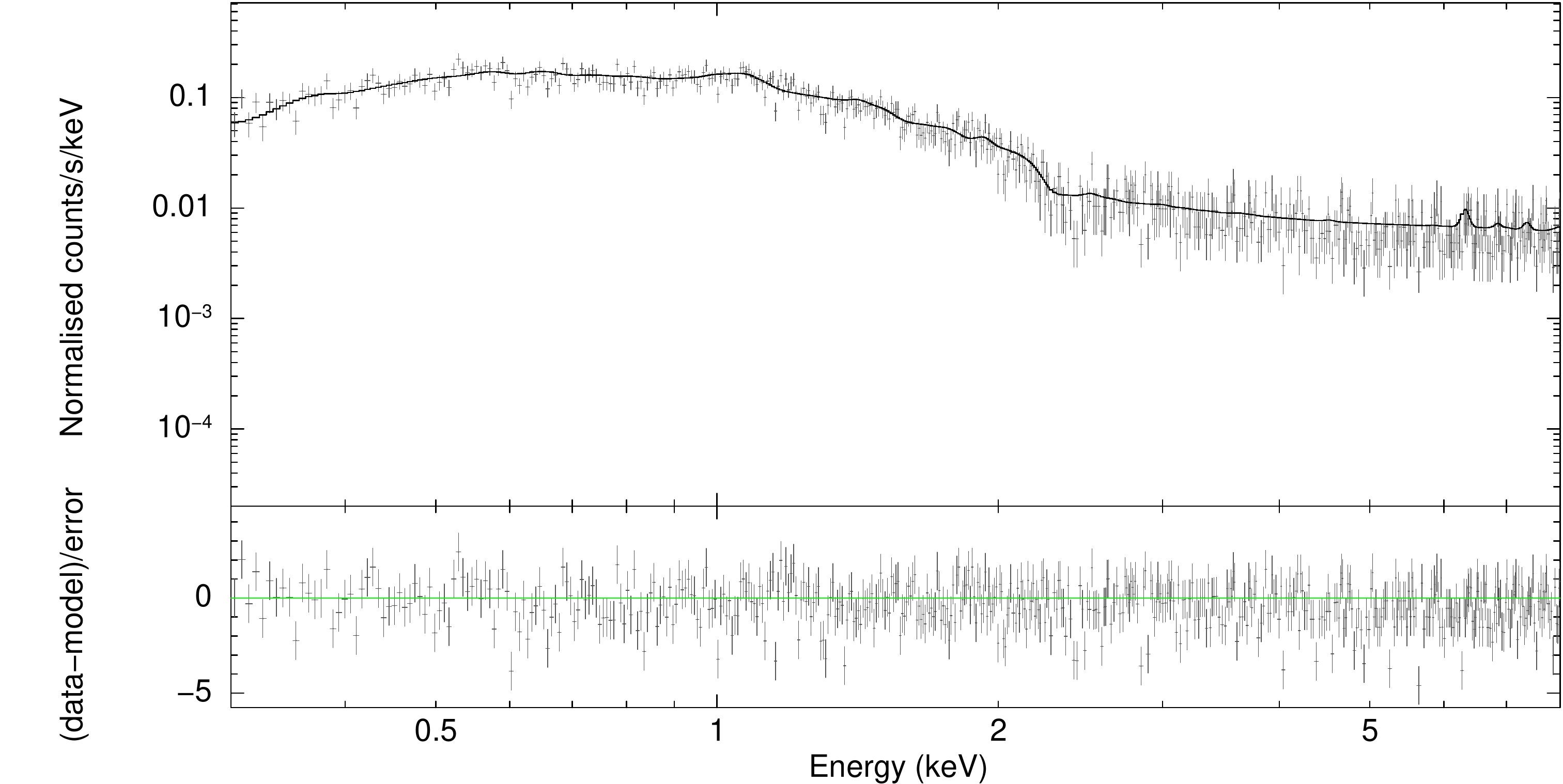}
	\caption{eROSITA spectrum of TM6 in $0.2r_{500}$ to $0.5r_{500}$ with corresponding model. 
		\label{fig:7}}
\end{figure}

In this procedure, the central region of A3408 is excluded to avoid emission from non-gravitational heating and cooling processes in the cluster center, i.e. the temperature is described by the depth of the gravitational potential \citep{Lovisari_2015}. After extracting the spectra from an arbitrary annulus, the spectral fitting is performed as previously explained. Fig.~\ref{fig:7} illustrates the spectrum and model of TM 6.
The mass estimation is then performed utilizing the $M_{500}-T$ relation 
\begin{align}
    \log(M_{500}/ c_1) = a \times \log(T/c_2) + b
\end{align}
from \cite{Lovisari_2015}, where $c_1=5 \times 10^{13}h_{70}^{-1}\:M_{\odot}$, $c_2=2\:{\rm keV}$, $a=(1.65 \pm 0.07)$, $b=(0.19 \pm 0.02)$ and $\log(x)$ is base 10. Assuming spherical symmetry, the characteristic radius $r_{500}$
can be derived from the estimated mass $M_{500}$.
We adopt a critical density of $\rho_{\rm crit}= 9.567 \times 10^{-27}\:{\rm kg}/{\rm m}^3$ at the cluster redshift $z_{\rm A3408}=0.042$ \footnote{\url{https://cosmocalc.icrar.org/}}.
Comparing the calculated $r_{500}$ with the source extraction area, one can iteratively find the temperature $T$ and mass $M_{500}$ within $0.2r_{500}$ to $0.5r_{500}$.

Iterating this procedure for A3408, it results in an ICM temperature of ${\rm k_B}T= 2.23 \pm 0.09 \:{\rm keV}$ and metallicity of $Z= 0.22 \pm 0.03 \:Z_{\odot}$ within an annulus of $2.8'$ to $7.3'$, corresponding to $0.2r_{500}$ to $0.5r_{500}$ respectively, where $r_{500}=(13.65 \pm 0.37)'$. 

As described before, $r_{200}$ and $r_{100}$ can be estimated from $r_{500}$. The results are listed in Tab.~\ref{tbl:2}. 


\begin{table}[h!]
	\centering
	\begin{tabular}[c]{ccccc}
		\hline
		${\rm k_B}T$ & $M_{500}$ & $r_{500}$ & $r_{200}$ & $r_{100}$ \\
		 $[{\rm keV}]$ & $[10^{13}M_{\odot}]$ & $[{\rm kpc}]$ & $[{\rm kpc}]$ & $[{\rm kpc}]$ \\
		\hline\hline
		& & & & \\
		$2.23 \pm 0.09$ & $9.27 \pm 0.75$ & $679 \pm 18$ & $1045 \pm 28$ & $1421 \pm 38$ \\
		& & & & \\
		\hline
	\end{tabular}
	\caption{Estimated cluster properties
	of A3408 applying the $M_{500}-T$ relation from \cite{Lovisari_2015}. The ICM temperature ${\rm k_B}T$
	is obtained from eROSITA spectral analysis within $0.2r_{500}$ to $0.5r_{500}$.
	The procedure is described in Sec.~\ref{sec:7.2}. 
	}
	\label{tbl:2}
\end{table}

Furthermore, we perform spectral fitting within a radius of $800\:{\rm kpc}$ as used in \cite{Katayama_2001} to compare with their results obtained from ASCA data. We find the best fit temperature ${\rm k_B}T= 1.93^{+0.05}_{-0.04}\:{\rm keV}$ and metallicity $Z= 0.20^{+0.02}_{-0.01}\:Z_{\odot}$. However, \cite{Katayama_2001} measured a temperature of ${\rm k_B}T=2.9^{+0.2}_{-0.2} \:{\rm keV}$ and metallicity of $Z= 0.34^{+0.20}_{-0.17}\:Z_{\odot}$ applying a single-temperature thermal emission model from \cite{Raymond_1977} in $0.7-10.0\:{\rm keV}$, while redshift $z$ and equivalent hydrogen column $N_H$ were equal to the values used in this work.
One possible reason for this deviation is the multitemperature structure of the ICM. There might
be gas present at different temperatures along the line of sight in the extraction area. If one uses a
single-temperature model to describe the spectra, the result depends on the observing telescope and its
response. A higher sensitivity at lower energies would result in stronger weighting of the Fe L shell emission line complex resulting in a lower temperature,
while a telescope with a higher sensitivity at higher energies would lead to a higher temperature, as
illustrated, e.g., in detail in \cite{Reiprich_2013}. eROSITA has more effective area at lower X-ray
energies compared to ASCA \citep{Predehl_2021,Tanaka_1994}, i.e. the multitemperature structure could qualitatively explain the lower
resulting temperature from eROSITA.

\subsection{Temperature, Metallicity and Normalization Profiles}
\label{sec:7.3}

Since A3408 has an elongated morphology, temperature, metallicity and normalization profiles are obtained from elliptical annuli out to a semi-major axis of $19'$ and a semi-minor axis of $9.5'$ to ensure enough source counts. The annuli have the same eccentricity and angle of $40\:{\rm deg}$, chosen visually aligned to the cluster shape, as shown in Fig.~\ref{fig:4}, center. 

\begin{figure}
\centering
	\includegraphics[width=1.0\columnwidth]{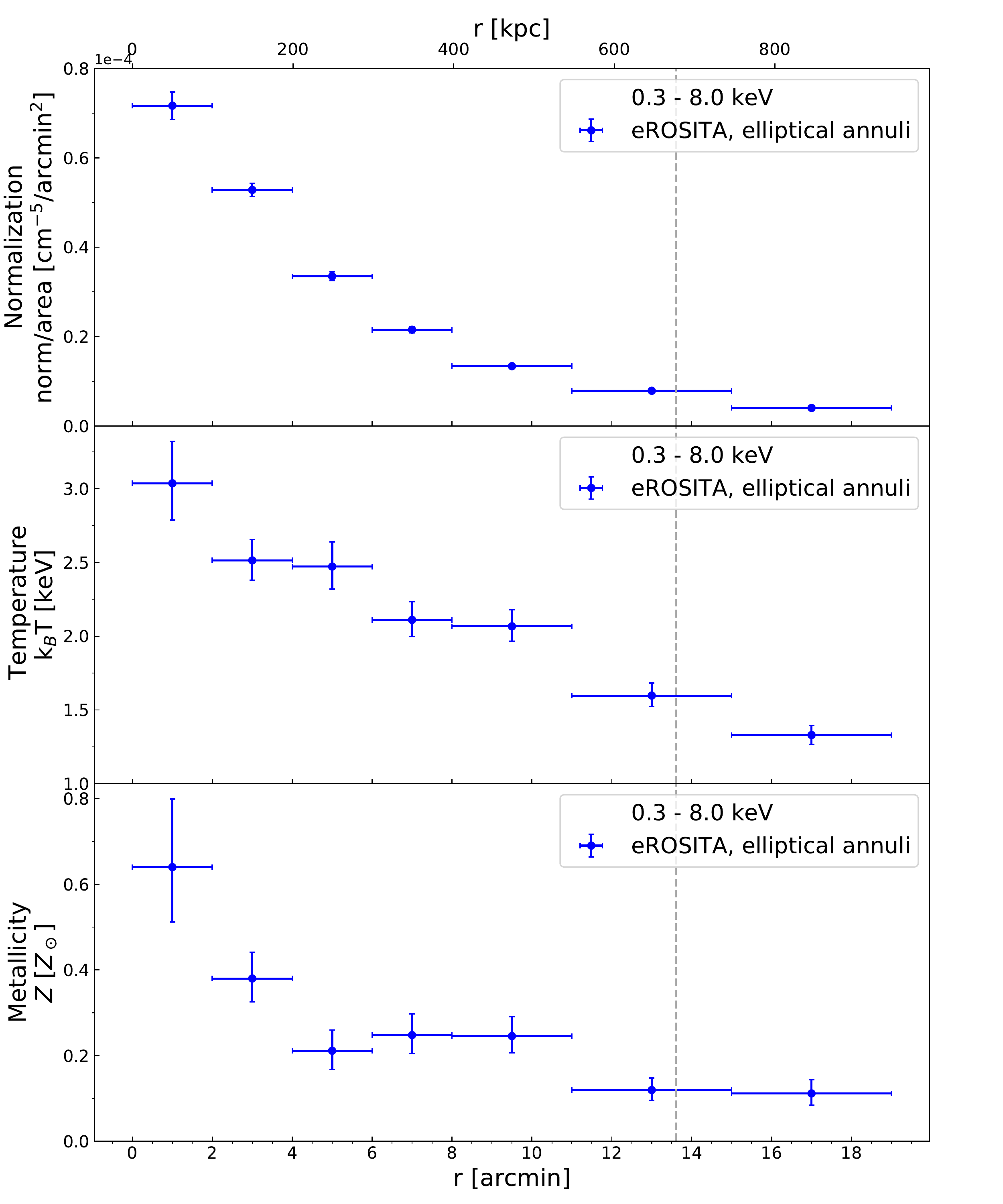}
	\caption{Normalization (top), temperature (middle), and metallicity (bottom) profiles of A3408 using elliptical annuli, where $r$ represents the semi-major axis.
	The grey dashed lines indicate the $r_{500}$ of the cluster.
		\label{fig:5}}
\end{figure}

The spectral fitting is performed in $0.3-8.0 \:{\rm keV}$ energy band for TM1, 2, and 6, and $1.0-8.0 \:{\rm keV}$ for TM 5 and 7
as previously explained. Resulting temperature, metallicity and normalization profiles are presented in Fig.~\ref{fig:5}.
We also performed spectral fitting of the NW half and SE half of the full elliptical annuli, subdivided along the semi-minor axis. The temperatures of the five outer regions of the profile show an indication for a systematic trend, i.e. each of them is higher in NW and lower in SE direction at $\sim$$1\sigma$ significance compared to the results of full annuli.

For further investigations of these differences and the temperature distribution of A3408, it is useful to create a temperature map with more bins aligned to the cluster emission.

\subsection{Temperature Map}
\label{sec:7.4}

In order to create a temperature map of the A3408 cluster, the region outside $19'$ is masked, along with point sources in the field, and the \cite{Sanders_2006} contour binning software is utilised to create a bin map of the cluster. 
The input parameters of the software determine the size and shape of the bins in the resulting bin map, where the distribution of the bins follows the surface brightness of the cluster. We apply the contour binning software to the combined image of TM 1, 2, 5, 6 and 7 in soft band to create a bin map with a number of counts of 2500 per bin.

Following the binning of the data, masks of each bin are created and used as input for the eSASS task \texttt{srctool} to perform spectral extraction for each TM. The spectral fitting is performed as previously described in Sec.~\ref{sec:7}. The resulting best fit temperature values and errors are then saved to the corresponding bin in the temperature map.

\begin{figure*}
\centering
	\includegraphics[width=2.05\columnwidth]{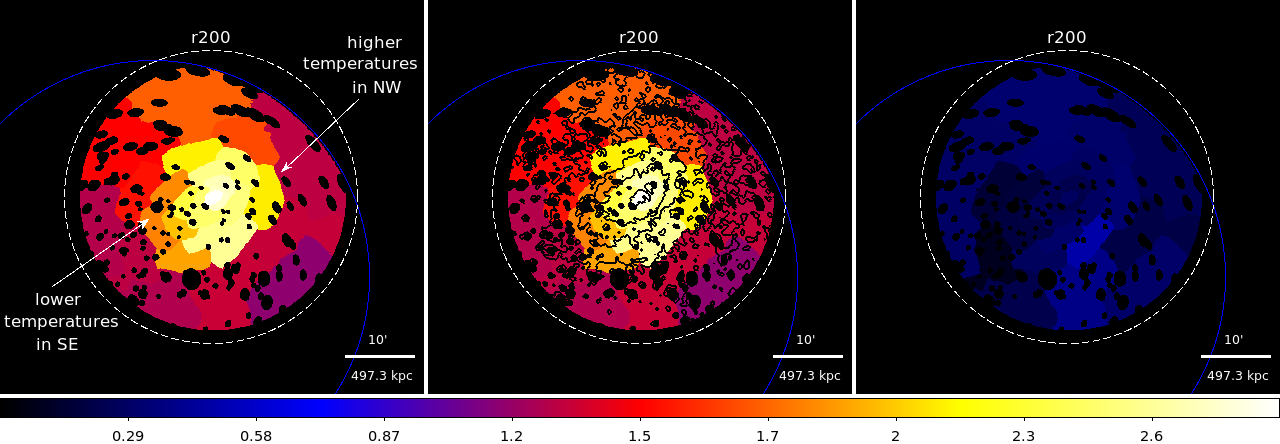}
	\caption{Temperature map of A3408. The eROSITA FoV (blue) and $r_{200}$ (white, dashed) are indicated.
	\textit{Left:} Temperature map with additional annulus of $4'-12'$ (black) for visualization. 
	\textit{Center:} Temperature map with surface brightness contours generated from the particle background subtracted and exposure corrected soft band image
	overlaid in black. 
	\textit{Right:} Relative errors for each temperature in corresponding bin. 
		\label{fig:6}}
\end{figure*}

The resulting temperature map with corresponding relative errors for each bin is shown in Fig.~\ref{fig:6}. Due to the elliptical morphology, as described in the imaging analysis, the contour binning method creates bins aligned to the elongation of A3408. One can see indeed a few more cooler bins in SE compared to NW. Furthermore, there seem to be higher temperatures in W compared to E. Relative errors are around $10\%-30\%$.

\section{Discussion}
\label{sec:8}

The imaging analysis underlines the intriguing morphology of A3408 elongated in SE-NW direction which is confirmed quantitatively in the comparison of the surface brightness profiles of four sectors NW, SW, SE and NE. This indicates that A3408 is not a relaxed system.
The ratio between the semi-major axis and semi-minor axis is 2.0.
The adaptively smoothed image shows enhanced emission towards A3407. Apart from this qualitative evidence, the surface brightness profile provides quantitative evidence of enhanced emission in the NW sector compared to the other regions. The relative differences between NW to SE are 60\% for the prominent peak at $15.3'$ 
(${\rm SB}_{{\rm SE}, 15.3'}=(2.56 \pm 0.54)\:10^{-7}{\rm counts/s/arcmin^2}$, 
${\rm SB}_{{\rm NW}, 15.3'}=(6.40 \pm 0.99)\:10^{-7}{\rm counts/s/arcmin^2}$), 68\% at $16.3'$ 
(${\rm SB}_{{\rm SE}, 16.3'}=(1.92 \pm 0.50)\:10^{-7}{\rm counts/s/arcmin^2}$, 
${\rm SB}_{{\rm NW}, 16.3'}=(5.98 \pm 1.22)\:10^{-7}{\rm counts/s/arcmin^2}$) and 79\% at $17.0'$ 
(${\rm SB}_{{\rm SE}, 17.0'}=(1.31 \pm 0.50)\:10^{-7}{\rm counts/s/arcmin^2}$, 
${\rm SB}_{{\rm NW}, 17.0'}=(6.17 \pm 1.44)$ $10^{-7}{\rm counts/s/arcmin^2}$). 

In fact, \cite{Nascimento_2016} describe a weak bridge  connecting A3407 and A3408 in a ROSAT PSPC image. However, they also mention the gap introduced by the support structure of the PSPC entrance window which could distort the appearance.
Nevertheless, the overlapping $r_{200}$ of both clusters, at least in projection, obtained from our ROSAT analysis of both clusters support the idea of A3407 and A3408 interacting which each other. 

Furthermore, the cluster properties $r_{500}, r_{200}, r_{100}$ are consistent with the results of \cite{Nascimento_2016}.
In their dynamical analysis of 122 member galaxies of A3407 and A3408, they found the galaxy velocity dispersions $\sigma_{\rm A3407} = 718^{+93}_{-48}\:{\rm km/s}$ and $\sigma_{\rm A3408} = 573^{+82}_{-59}\:{\rm km/s}$ and the corresponding virial masses $\log(M_{\rm V})_{\rm A3407} = 14.59^{+0.46}_{-0.42}\:h^{-1}M_{\odot}$ and $\log(M_{\rm V})_{\rm A3408} = 14.26^{+0.55}_{-0.46}\:h^{-1}M_{\odot}$. With these results and the assumption $M_{\rm V} \approx M_{100}$, the characteristic radii $r_{100}, r_{200}, r_{500}$ can be roughly estimated again with the scaling relations from \cite{Reiprich_2013}. One finds $r_{100} \approx 33', r_{200} \approx 24'$ and  $r_{500} \approx 16'$ for A3408 
which is in agreement with the findings in this work. For A3407 one finds $r_{100} \approx 42', r_{200} \approx 31'$ and  $r_{500} \approx 20'$
which is significantly larger than the results from analysis of the ROSAT data. However, velocity dispersions for close systems as A3408 and A3407 might result in overestimated mass estimates. This might explain the discrepancy for A3407. From the X-ray images; i.e., from the X-ray luminosities, it appears that A3408 should be the more massive cluster and not A3407 as \cite{Nascimento_2016} found. 

In the temperature map, as well as in the temperature profile, a hot core is discovered.
Furthermore, one can see a few lower temperatures in southern bins compared to the northern bins but also that the west seems to be hotter than the east. A3407 lies in the NW direction, therefore, this could indicate a shock front or adiabatically compressed gas due to onsetting interaction with A3407. How such early interaction could lead to a hot core in A3408 is not completely obvious; possibly, it is the result of a recent previous merger, which could also have contributed to forming the very elongated shape.


The outer three bins in W (in $4'-12'$) have temperatures of $T_{{\rm W},1}=2.60^{+0.82}_{-0.42}~{\rm keV}$, $T_{{\rm W},2}=2.13^{+0.35}_{-0.31}~{\rm keV}$ and $T_{{\rm W},3}=2.47^{+0.37}_{-0.32}~{\rm keV}$. While in E (in $4'-12'$) we find temperatures of $T_{{\rm 
E},1}=1.91^{+0.29}_{-0.21}~{\rm keV}$, $T_{{\rm E},2}=1.83^{+0.19}_{-0.15}~{\rm keV}$ and $T_{{\rm E},3}=1.85^{+0.16}_{-0.17}~{\rm keV}$. 
The discrepancies of the temperatures are at $2\sigma$ for $T_{{\rm W},1}$ and $T_{{\rm 
E},1}$, $1\sigma$ for $T_{{\rm 
W},2}$ and $T_{{\rm 
E},2}$ and $3\sigma$ for $T_{{\rm 
W},3}$ and $T_{{\rm 
E},3}$.

\section{Conclusions}
\label{sec:9}

The galaxy cluster A3408 has been serendipitously observed during eROSITA's Performance Verification phase.
The imaging analysis underlines the intriguing elliptical cluster morphology in SE-NW direction. 
We determine for A3408 a gas temperature of ${\rm k_B}T_{500}=(2.23\pm0.09)\:{\rm keV}$ within $0.2r_{500}$ to $0.5r_{500}$, a mass of $M_{500} = (9.27 \pm 0.75)\times 10^{13}M_{\odot}$ and $r_{500} = (679 \pm 18)\:{\rm kpc}$.
There are indications that A3407 and A3408 are interacting, e.g. enhanced emission from A3408 towards A3407 in the adaptively smoothed image and the surface brightness profile, and enhanced temperatures in the NW direction of A3408.
We interpret this such that A3408's ICM thermodynamic properties are affected by early interaction with the A3407 cluster. Hence, this cluster pair is another very good candidate for studying bridge emission in X-rays and radio -- as the systems A399/401 \citep[e.g.,][]{Planck_2013} and A3391/95 \citep[e.g.,][]{Reiprich_2020,2021A&A...647A...3B} -- in order to understand hydrodynamical processes including the early onset of turbulence.

The eROSITA All-Sky-Survey (eRASS) will enable the analysis of the whole A3407-A3408 system. Cluster properties of A3407 together with this analysis will lead to a better understanding of this system. Furthermore, if there is an emission bridge connecting both clusters, the eRASS image will provide its observation.

\begin{acknowledgements}
\\
Part of this work has been funded by the Deutsche Forschungsgemeinschaft (DFG, German Research Foundation) – 450861021.

This work is based on data from eROSITA, the soft X-ray instrument aboard SRG,
a joint Russian-German science mission supported by the Russian Space Agency (Roskosmos), in the interests of the Russian Academy of Sciences represented by its Space Research Institute (IKI), and the Deutsches Zentrum für Luft- und Raumfahrt (DLR). The SRG spacecraft was built by Lavochkin Association (NPOL) and its subcontractors, and is operated by NPOL with support from the Max Planck Institute for Extraterrestrial Physics (MPE).\\

The development and construction of the eROSITA X-ray instrument was led by MPE, with contributions from the Dr. Karl Remeis Observatory Bamberg \& ECAP (FAU Erlangen-Nuernberg), the University of Hamburg Observatory, the Leibniz Institute for Astrophysics Potsdam (AIP), and the Institute for Astronomy and Astrophysics of the University of Tübingen, with the support of DLR and the Max Planck Society. The Argelander Institute for Astronomy of the University of Bonn and the Ludwig Maximilians Universität Munich also participated in the science preparation for eROSITA.\\

The eROSITA data shown here were processed using the eSASS/NRTA software system developed by the German eROSITA consortium.

\end{acknowledgements}

\bibliographystyle{aa} 
\bibliography{bib.bib}

\begin{appendix}

\section{Lightcurves}
\label{A1}

\begin{figure}[h!]
\centering
	\includegraphics[width=0.9\columnwidth]{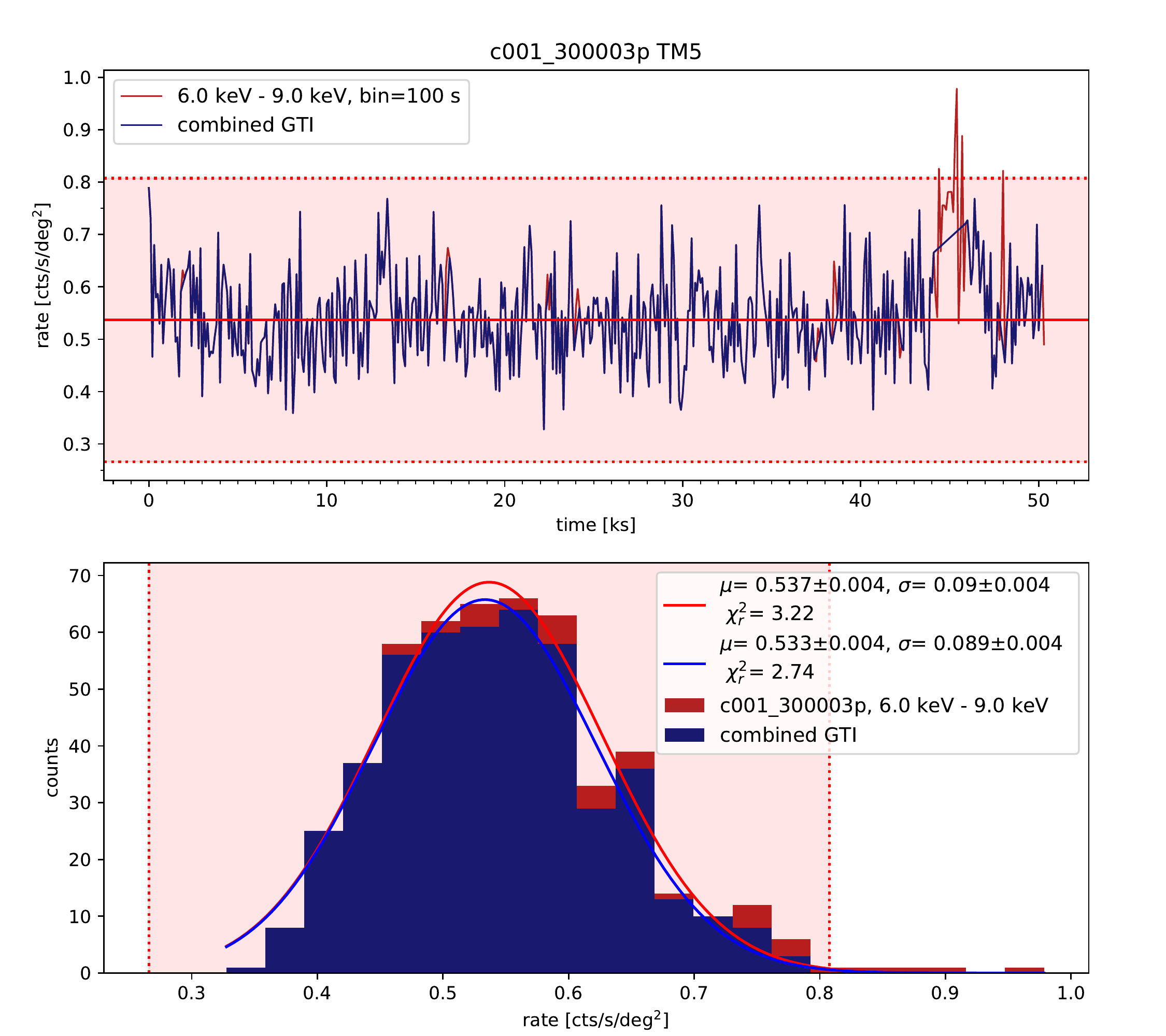}
	\includegraphics[width=0.9\columnwidth]{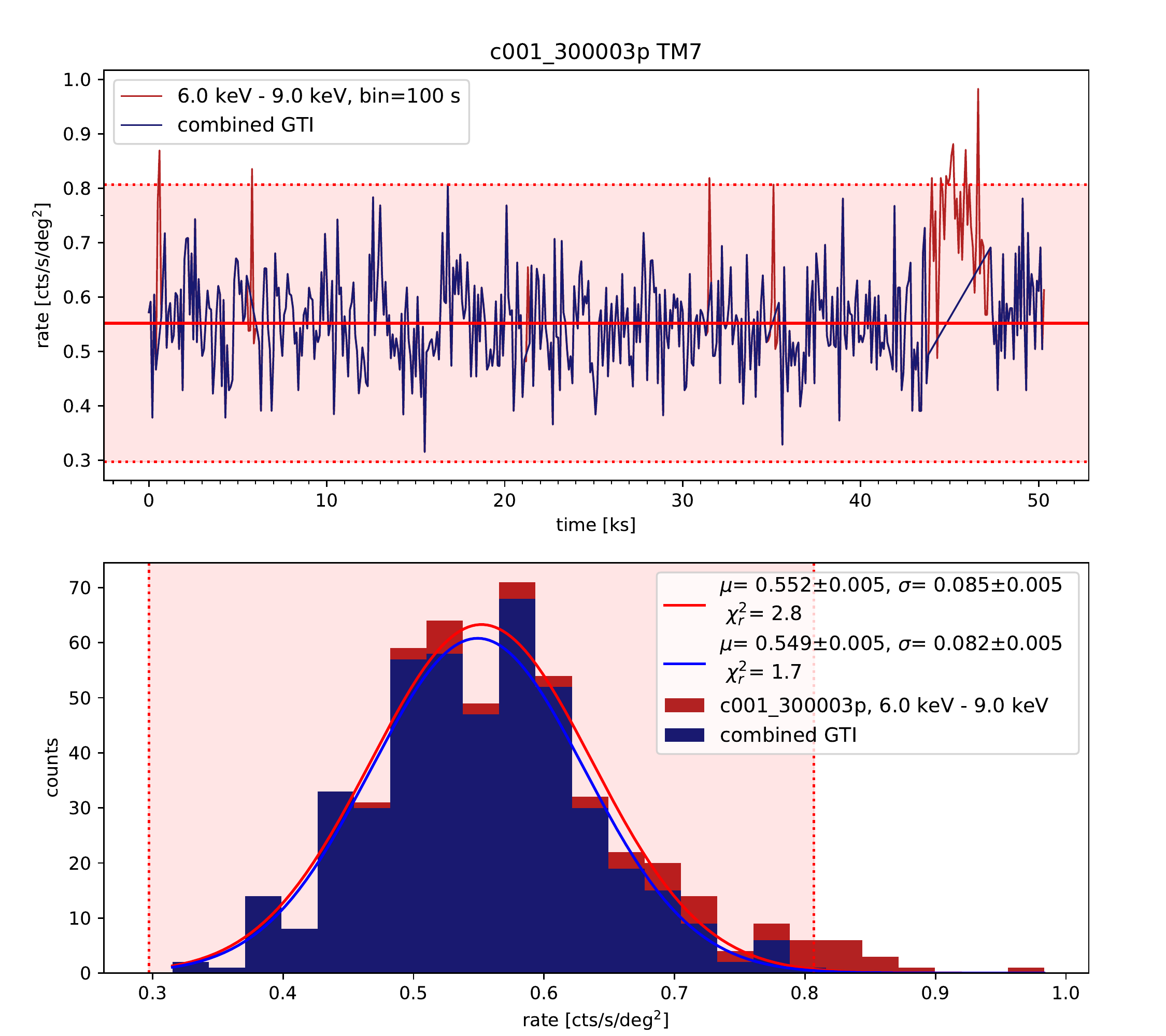}
	\caption{Lightcurves of TM5 and TM7 (without on-chip filter) filtered from events outside $3\sigma$-interval (red-shaded area).
		\label{fig:A1}}
\end{figure}

\begin{figure}
\centering
	\includegraphics[width=0.9\columnwidth]{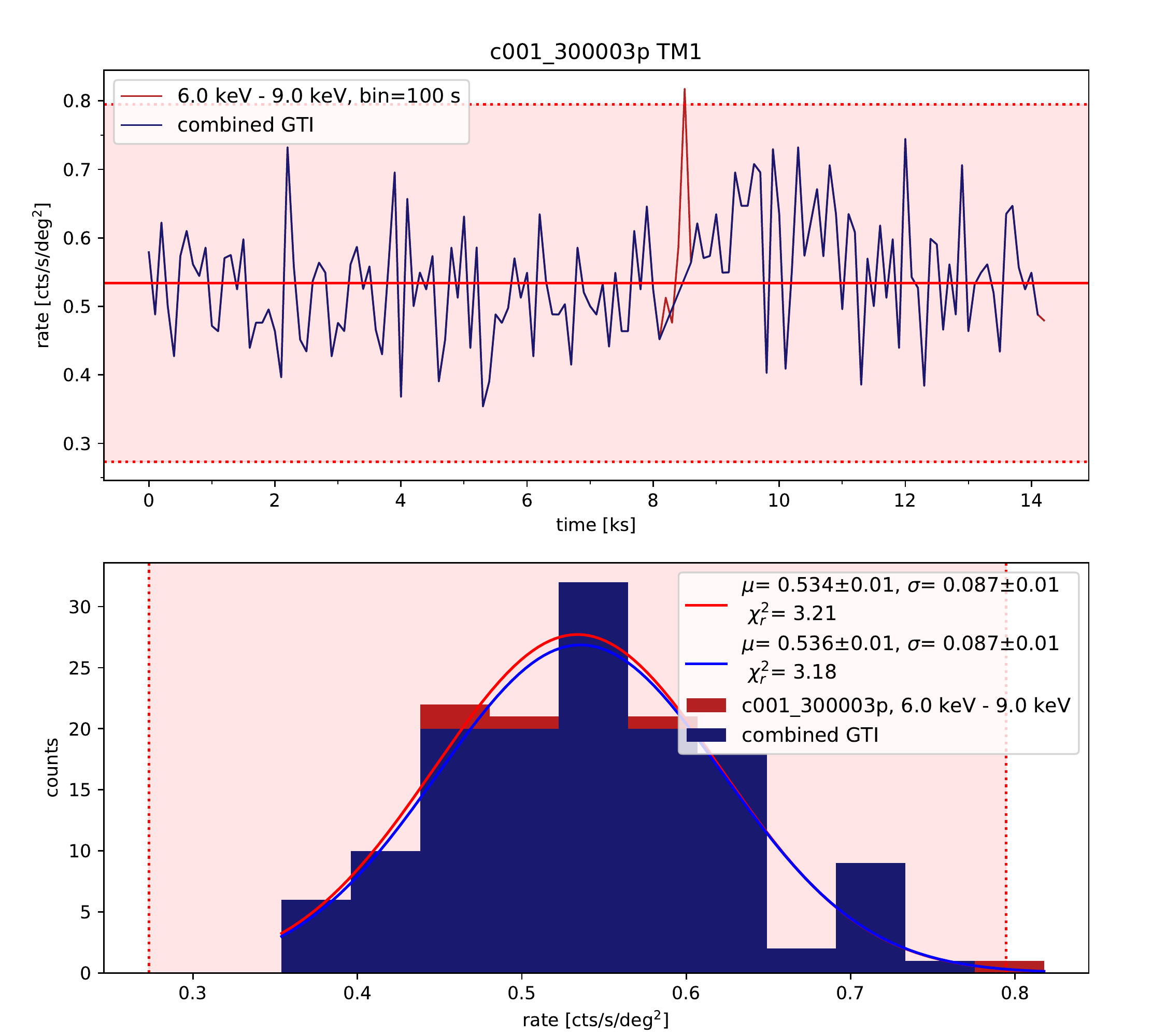}
	\includegraphics[width=0.9\columnwidth]{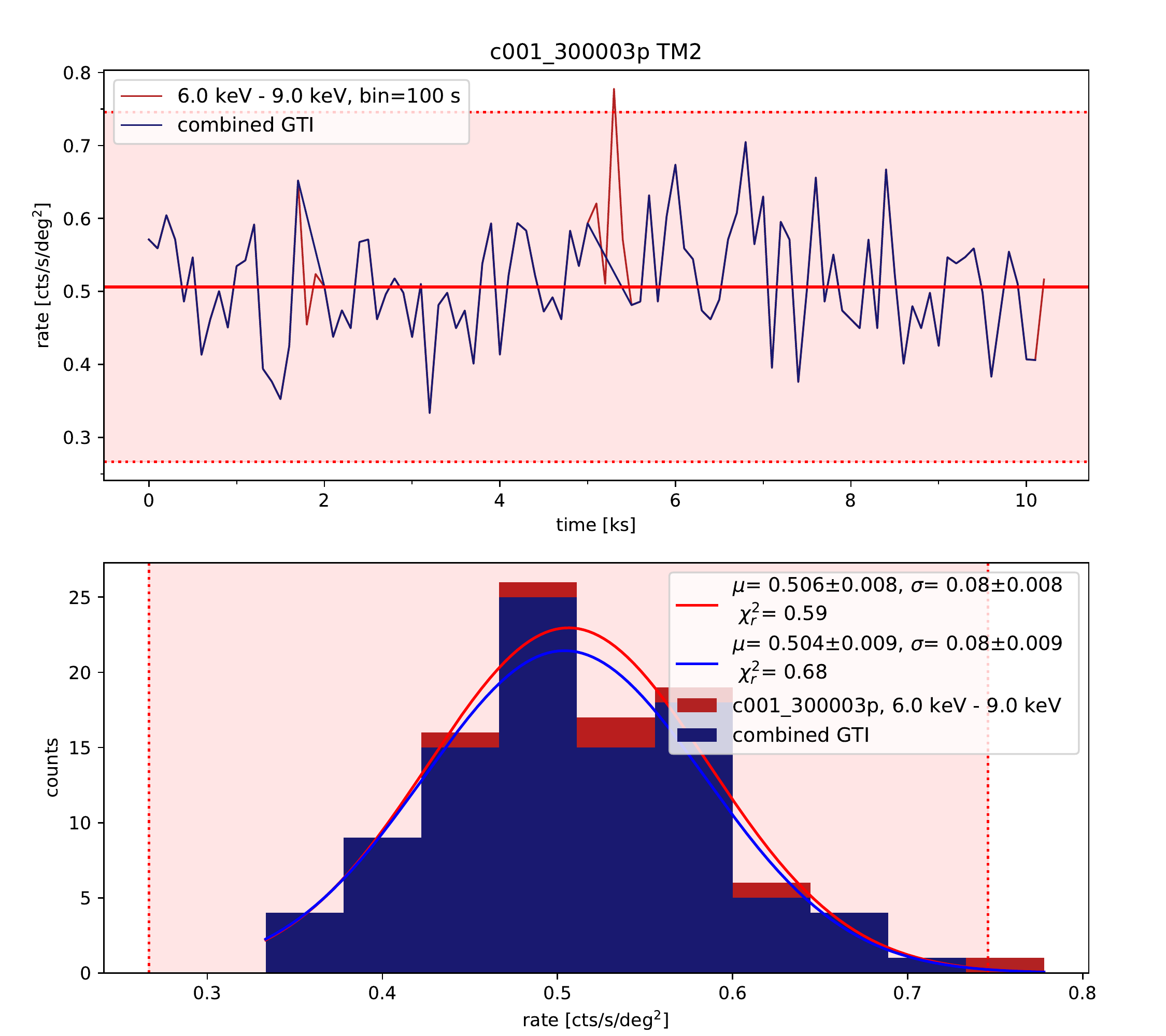}
	\includegraphics[width=0.9\columnwidth]{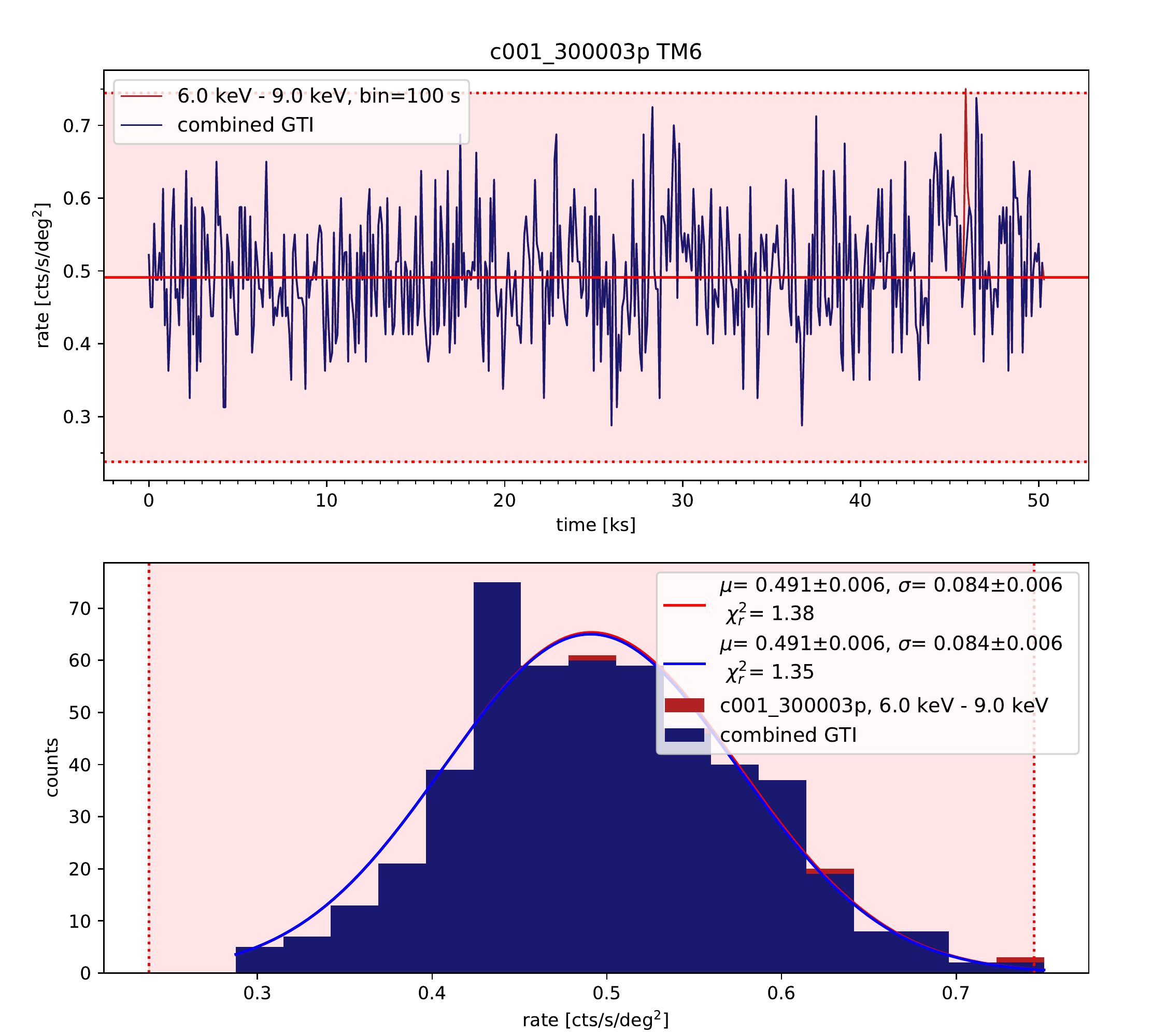}
	\caption{Lightcurves of TM1, TM2 and TM6 (with on-chip filter) filtered from events outside $3\sigma$-interval (red-shaded area).
		\label{fig:A2}}
\end{figure}

Fig.~\ref{fig:A1} shows generated lightcurves of TM without on-chip filter and Fig.~\ref{fig:A2} with on-chip filter using the eSASS task \texttt{flaregti} in the energy band of 6.0~keV to 9.0~keV. All events outside the $3\sigma$-interval represented by the red-shaded area are excluded. 

In the top panels, the lightcurves of the whole observations (red curves) and the lightcurves after the filtering procedure (blue curves) are plotted. The filtering procedure includes the automatic and additional filtering steps. The automatic filtering results in Good-Time Intervals (GTIs) generated by the standard eSASS processing. The additional filtering is done by determining the $3\sigma$ interval by fitting a Gaussian model to the histogram of the rates (bottom panels). Any time intervals beyond $\pm3\sigma$ (red-shaded areas) are rejected from the data. For instance, the excluded red spikes within $3\sigma$ are from the automatic filtering
and those outside $3\sigma$ are from additional filtering. Finally, the GTIs from both automatic and additional filtering are combined and used for the next steps.

\end{appendix}

\end{document}